\documentclass[runningheads]{llncs}
\usepackage[T1]{fontenc}
\usepackage{graphicx}
\usepackage{amsmath,amssymb}
\usepackage{booktabs}
\usepackage{subcaption}
\usepackage{tikz}
\usepackage{tabularx}
\usepackage{url}
\usepackage{rotating}
\usepackage{subcaption}
\usepackage{xcolor}

\begin{document}

\title{Using Continual Learning for Real-Time Detection of Vulnerable Road Users in Complex Traffic Scenarios}

\author{Faryal Aurooj Nasir\inst{1} \and Salman Liaquat\inst{2} \and Nor Muzlifah Mahyuddin\inst{2}\thanks{Corresponding Author: eemnmuzlifah@usm.my}}

\authorrunning{F.A. Nasir et al.}

\institute{
Department of Electrical Engineering, Institute of Space Technology, 44000 Islamabad, Pakistan 
\and
School of Electrical and Electronic Engineering, Universiti Sains Malaysia, 14300 Penang, Malaysia
}

\maketitle

\begin{abstract}
Pedestrians and bicyclists are among the vulnerable road users (VRUs) that are inherently exposed to intricate traffic scenarios, which puts them at increased risk of sustaining injuries or facing fatal outcomes. This study presents an intelligent adaptive system that uses the YOLOv8-Dynamic (YOLOv8-D) algorithm that detects vulnerable road users and adapts in real time to prevent accidents before they occur. We select YOLOv8x as the detector by comparing it with other state-of-the-art object detection models, including Faster-RCNN, YOLOv5, YOLOv7, and variants. Compared to YOLOv5x, YOLOv8x shows improvements of 12.14\% in F1 score and 45.61\% in mean Average Precision (mAP). Against YOLOv7x, the improvements are 21.26\% in F1 score and 128.44\% in mAP. Our algorithm integrates continual learning ability in the architecture of the YOLOv8 detector to adjust to evolving road conditions flexibly, ensuring adaptability across multiple dataset domains and facilitating continuous enhancement of detection and tracking accuracy for VRUs, embracing the dynamic nature of real-world environments. In our proposed framework, we optimized the gradient descent mechanism of YOLOv8 model and train our optimized algorithm on two statistically different datasets in terms of image viewpoint and number of classes to achieve a 21.08\% improvement in F1 score and a 31.86\% improvement in mAP as compared to a custom YOLOv8 framework trained on a new dataset, thus overcoming the issue of catastrophic forgetting, which occurs when deep models are trained on statistically different types of datasets.



\end{abstract}
\section{Introduction}

Traffic collisions result from various factors, including reckless driving, speeding, impaired driving, failure to obey signals, and dangerous manoeuvres. Speed, in particular, is a significant contributor to road traffic injuries \cite{gatera2022towards}. According to the WHO's Global Status Report on Road Safety 2018, road traffic fatalities in 2016 were 1.35 million \cite{WHO2018RoadSafety}, and while the number slightly decreased to 1.19 million in 2023, road traffic accidents remain the leading cause of death for individuals aged 5 to 29, with over 3,200 fatalities daily \cite{WHO2023RoadSafety}. Vulnerable groups, such as pedestrians, motorcyclists, and cyclists, are particularly affected. The economic burden of road traffic injuries, particularly fatalities, is substantial, placing strain on national economies. Although complete elimination of traffic accidents is not feasible, efforts can be made to reduce their occurrence and severity. The United Nations has set a goal to reduce road traffic fatalities by 50\% by 2030 as part of the Sustainable Development Agenda (Goal 11.2) \cite{UN_SDG2030}.

Self-driving vehicles (SDVs), once a futuristic concept, have made significant progress in recent years. Detecting other road users is crucial for SDV safety, as efficient detection improves decision-making and prevents accidents \cite{djuric2020uncertainty}.
Road planners and policymakers have adjusted legal structures, road designs, and education for drivers and vulnerable road users (VRUs) \cite{constant2010protecting}, but these measures have had limited success in addressing SDV-related challenges. Advanced detection and tracking mechanisms are essential to improving road safety and contributing to the achievement of the WHO's goal of reducing traffic fatalities and injuries by 50\%.
Over the past decade, methodologies to detect VRUs and predict their motion trajectories have evolved significantly. Early techniques, such as the Hidden Markov Model, relied on statistical signal processing for detection probabilities. Since 2015, the focus has shifted to classical machine learning and modern deep learning techniques. Traditional methods, including wavelet transform \cite{elzein2003motion}, \cite{schauland2006vision}, hand-crafted features for pedestrian detection \cite{xu2011pedestrian}, and support vector machines, were constrained by assumptions like stationary cameras and frontal pedestrian views. Other approaches, such as combining radar and cameras \cite{dimitrievski2019people}, faced computational efficiency challenges.

Deep learning methods have outperformed traditional ones, especially with models like deep neural networks \cite{karakaya2023cyclesense}, recurrent neural networks (RNNs) \cite{islam2023traffic}, and convolutional neural networks (CNNs). Significant advancements in real-time object detection have been achieved with detectors like YOLO (You Only Look Once) \cite{redmon2016you} and Faster R-CNN \cite{zhao2016faster}. YOLO, particularly, is widely adopted for real-time object detection \cite{zhao2023yolov7}, with frameworks such as YOLOv4, YOLOv5, and YOLOv7 being used in dynamic environments for pedestrian detection and tracking \cite{wang2023detector}.
YOLOv5 uses a modified EfficientNet backbone \cite{tan2019efficientnet}, while YOLOv7 employs Efficient Layer Aggregation Network (ELAN) for better feature extraction. YOLOv8 improves upon YOLOv7 with mosaic data augmentation and an anchor-free detection system, enhancing generalisation and speeding up object detection. YOLOv8's architecture includes a C2f module and a decoupled head for superior classification and regression. Advanced loss functions like Complete IoU (CIoU) and Dynamic Focal Loss (DFL) refine bounding box predictions, improving accuracy. The Adam optimiser, with adaptive learning rates, aids in model convergence. Deep learning-based approaches, especially YOLO frameworks, significantly advance VRUs detection and tracking, essential for road safety and autonomous vehicle development.

Catastrophic forgetting, where models fail to retain prior information while learning new data, poses a challenge, particularly in real-time object detection. Continual learning addresses this by enabling models to learn from continuous data streams without forgetting previous knowledge \cite{kirkpatrick2017overcoming}, crucial for tasks requiring ongoing learning akin to human learning. Transfer learning adapts pre-trained models to new tasks \cite{torrey2010transfer,weiss2016survey}, while continual learning allows for continuous task updates without loss of prior information \cite{hassabis2017neuroscience}. Techniques like gradient-based methods, modularity-based methods, and knowledge distillation combat catastrophic forgetting, helping models adapt to new data while retaining existing knowledge \cite{arpit2017closer,lopez2017gradient,zenke2017continual,aljundi2018memory,hinton2015distilling,li2017learning}

In this paper, we propose the YOLOv8-Dynamic (YOLOv8-D) algorithm, which is a novel extension of YOLOv8 that incorporates continual learning capabilities, enabling the system to adapt to changing road conditions and continuously improve the VRU detection accuracy.  The key contributions of this study are as follows: 
\begin{itemize}    
\item The paper compares various detection frameworks, such as Faster R-CNN, YOLOv5, YOLOv7, YOLOv8, and their variants, for real-time VRU detection. After careful analysis, YOLOv8x was identified as the most suitable framework because of its superior performance in real-time VRU detection, demonstrating the strength of YOLOv8x in dynamic environments.  
\item The research extends YOLOv8x for continual learning, enabling the model to adapt to new and changing road environments over time. By optimizing the gradient descent method, the results demonstrate that the model can avoid catastrophic forgetting, ensuring that it retains the knowledge from earlier tasks while learning from new data. 
\end{itemize}

These contributions address significant challenges in real-time VRU detection and continual learning with a focus on enhancing the adaptability and performance of autonomous vehicle systems in dynamic traffic environments. The paper is structured as follows: Section \ref{proposed} presents the detailed methodology, including the selection of the detection model and integration of continual learning, followed by Section \ref{results} which discusses the experimental results and analysis. Finally, Section \ref{conclusion} concludes the paper.

\section{Proposed Methodology} \label{proposed}
The proposed methodology for detecting VRUs in dynamic environments integrates multiple advanced techniques, including dataset selection and processing, model selection, and continual learning incorporation.

\subsection{Datasets Selection}
This study utilizes primarily two datasets for training and evaluating VRU detection and continual learning characteristics: VisDrone2019 \cite{zhu2021detection} dataset and Caltech Pedestrian dataset \cite{dollar_wojek_schiele_perona_2009}. We utilized 10,209 static images of the VisDrone2019 dataset for training, validation, and testing. The training dataset comprised 6000 images, the validation set comprised 2209 images, and the test set contained 2000 images. The Caltech pedestrian dataset includes 350,000 bounding boxes annotated for about 2300 distinct pedestrians. We trained our models on this dataset to evaluate the continual learning performance of the proposed algorithm by utilizing 5000 images for training, 1500 for validation and 500 for testing. This dataset is used for training the YOLOv8 model before and after changing the optimizer from Adam optimizer to Stochastic Gradient Descent (SGD) optimizer. 

\subsection{Dataset Processing}  
The VisDrone2019 dataset, originally annotated for over 9 classes, was re-annotated to focus on VRUs, specifically pedestrians, people standing by roadsides, bicyclists, and tricycles. The annotations were then converted into the YAML format required for training YOLO models, ensuring better inference speed and accuracy in cluttered environments. The modified dataset is suitable for training YOLOv5, YOLOv7, and YOLOv8 models specific to this task.

\subsection{Framework and Model Size Exploration} 
We systematically evaluated various object detection frameworks, including Faster R-CNN and YOLO models across multiple sizes (e.g., small and extra-large configurations) to identify the optimal balance between accuracy and inference speed. This exploration led to the selection of YOLOv8x, which demonstrated superior performance in real-time VRU detection, offering high accuracy while reducing inference latency.

\subsection{Continual Learning Incorporation} 
To address the challenge of catastrophic forgetting, we incorporated continual learning into the YOLOv8 model using gradient-based methods. The Adam optimizer was initially used, which offered adaptive learning rates, but to improve the generalization capability of the model and prevent the phenomenon of forgetting previously learned features in order to optimize on a new dataset, we changed the optimizer with the SGD algorithm. SGD's inherent randomness helps regularize the learning process and allows the model to adapt to new data while retaining knowledge from previously learned tasks. Our results demonstrate that this approach proved effective in incremental learning scenarios, especially when training on the Caltech Pedestrian dataset after the initial training on the VisDrone2019 dataset.

\subsection{Performance Evaluation}  
Multiple performance metrics were used to evaluate the detection effectiveness of the proposed system. To quantitatively analyze the performance of detection models, the following equations show the performance metrics such as Precision (P), Recall (R), mean Average Precision (mAP) for all classes, and F1 scores that have been used to evaluate the results,

\[ \text{Precision} = \frac{TP}{TP + FP}, \]
\vspace{-0.2cm}
\[ \text{Recall} = \frac{TP}{TP + FN}, \]
\vspace{-0.2cm}
\[ \text{mAP} = \frac{1}{N} \sum_{i=1}^{N} AP_i, \]
\vspace{-0.2cm}
\[ \text{F1 Score} = \frac{2 \times Precision \times Recall}{Precision + Recall}, \]
where

\begin{itemize}
    \item TP (True Positives): The number of correctly predicted positive instances by the model,
    
    \item FP (False Positives): The number of instances that were incorrectly predicted as positive by the model,
    
    \item FN (False Negatives): The number of positive instances that were incorrectly predicted as negative by the model, and

    \item AP is the average precision for each class.
    
\end{itemize}

\section{Results \& Discussion} \label{results}
\vspace{-0.2cm}
\subsection{Detection Model Evaluation}

\begin{sidewaystable}[htbp]
\caption{Performance metrics for VRU classes (Pedestrian, People, Tricycle, Bicycle) detected by Faster R-CNN, YOLOv5s, YOLOv5x, YOLOv7s, YOLOv7x, YOLOv8s, and YOLOv8x models.}\label{tab2}
\begin{tabular}{|l|l|l|l|l|l|l|l|}
\hline
\textbf{Metric} & \textbf{Faster R-CNN} & \textbf{YOLOv5s} & \textbf{YOLOv5x} & \textbf{YOLOv7s} & \textbf{YOLOv7x} & \textbf{YOLOv8s} & \textbf{YOLOv8x} \\
\hline
Precision @ 0.2 confidence & 0.55 & 0.455 & 0.582 & 0.545 & 0.571 & 0.568 & 0.763 \\
\hline
Recall @ 0.2 confidence & 0.67 & 0.292 & 0.381 & 0.342 & 0.215 & 0.335 & 0.485 \\
\hline
mAP @ 0.50 & 0.523 & 0.291 & 0.353 & 0.219 & 0.225 & 0.357 & 0.514 \\
\hline
F1 score @ 0.20 confidence & 0.601 & 0.343 & 0.412 & 0.341 & 0.381 & 0.414 & 0.462 \\
\hline
FPS & 4.55 & 175 & 24 & 290 & 46 & 625 & 101 \\
\hline
Inference Time (ms) & 220 & 5.7 & 41.1 & 12.6 & 21.5 & 1.6 & 9.9 \\
\hline
Training Time (@300 epochs) & 36.5 & 16.62 & 20.72 & 22.5 & 28.87 & 8.675 & 15.831 \\
\hline
Computational Time per 30 FPS Input (sec) & 6.59 & 0.17 & 1.25 & 0.10 & 0.65 & 0.048 & 0.297 \\
\hline
\end{tabular}
\end{sidewaystable}

\begin{figure}[htbp]
\centering

\begin{subfigure}[b]{0.49\textwidth}
    \includegraphics[width=\textwidth]{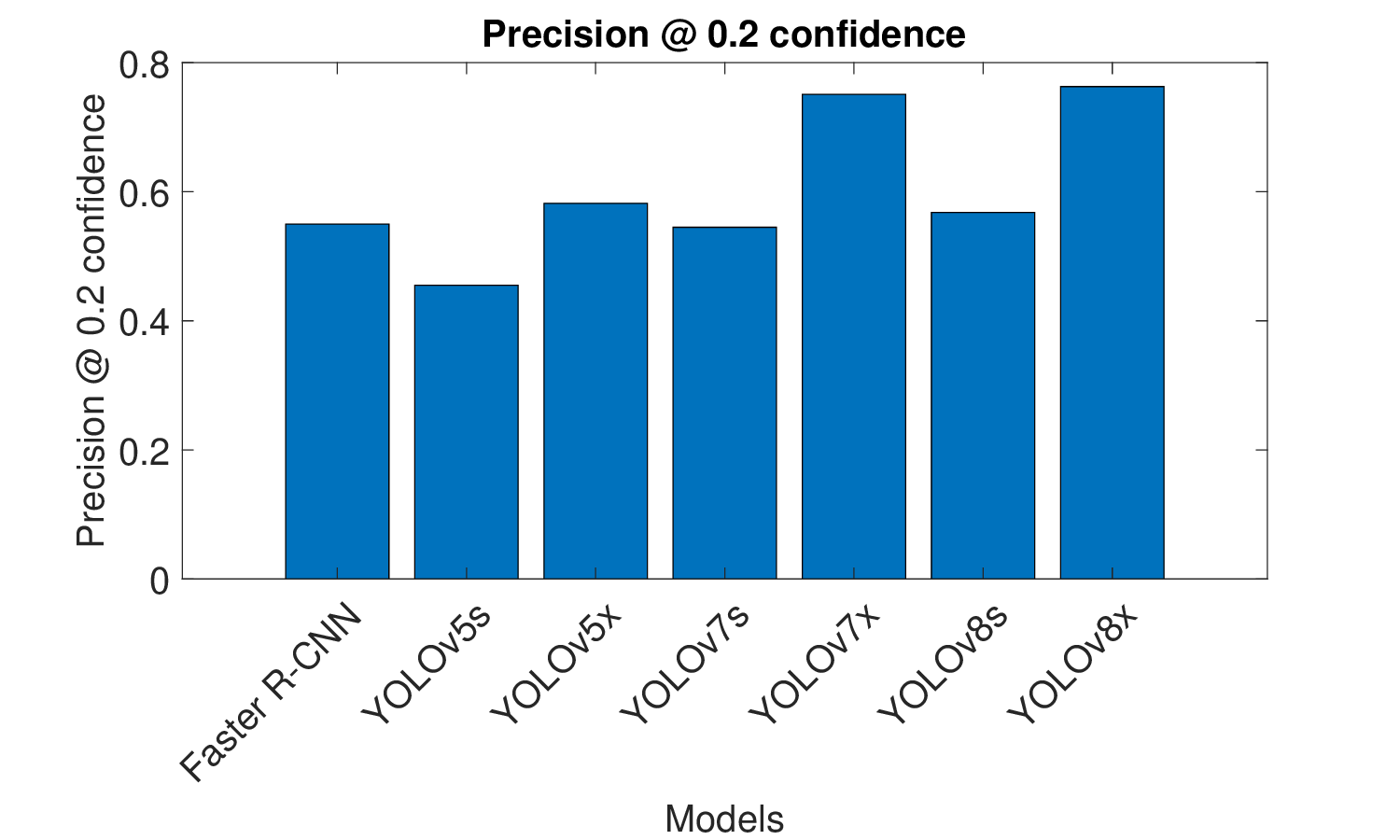}
    \caption{Comparison of precision (@0.2 confidence threshold).}
    \label{fig:precision}
\end{subfigure}
\hfill
\begin{subfigure}[b]{0.49\textwidth}
    \includegraphics[width=\textwidth]{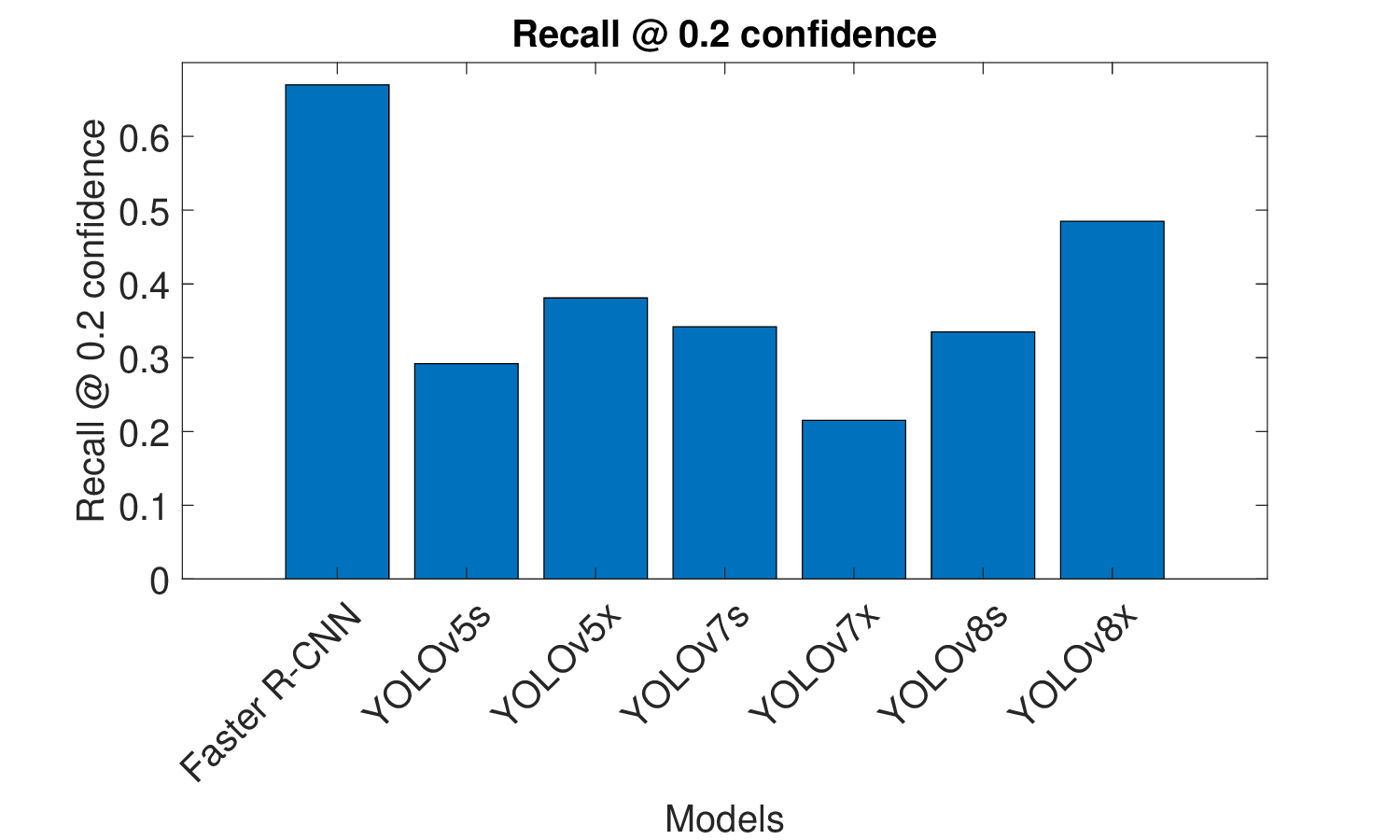}
    \caption{Comparison of recall (@0.2 confidence threshold).}
    \label{fig:recall}
\end{subfigure}
\vspace{0.5cm}

\begin{subfigure}[b]{0.49\textwidth}
    \includegraphics[width=\textwidth]{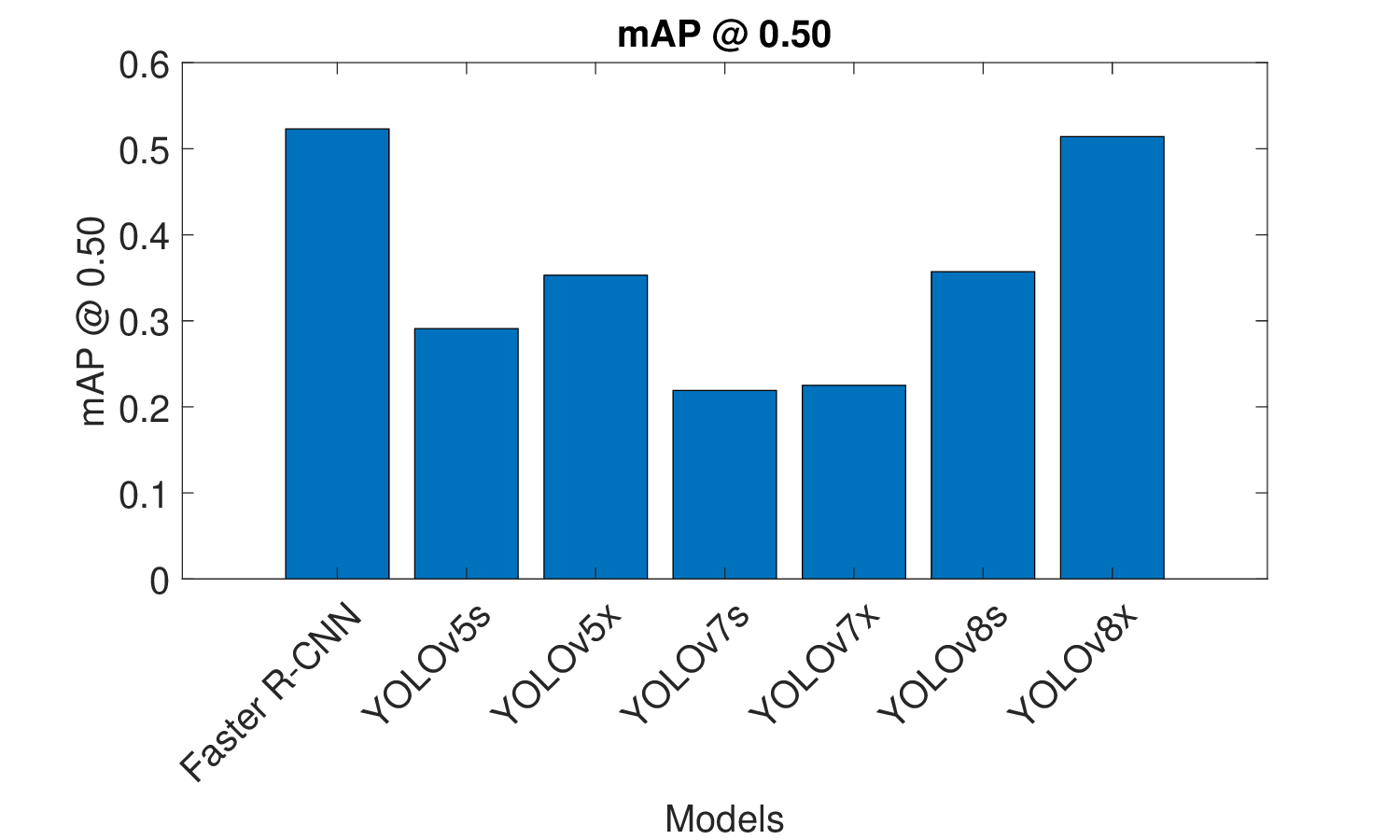}
    \caption{Comparison of mean average precision (mAP) @50 IoU threshold.}
    \label{fig:mAP}
\end{subfigure}
\hfill
\begin{subfigure}[b]{0.49\textwidth}
    \includegraphics[width=\textwidth]{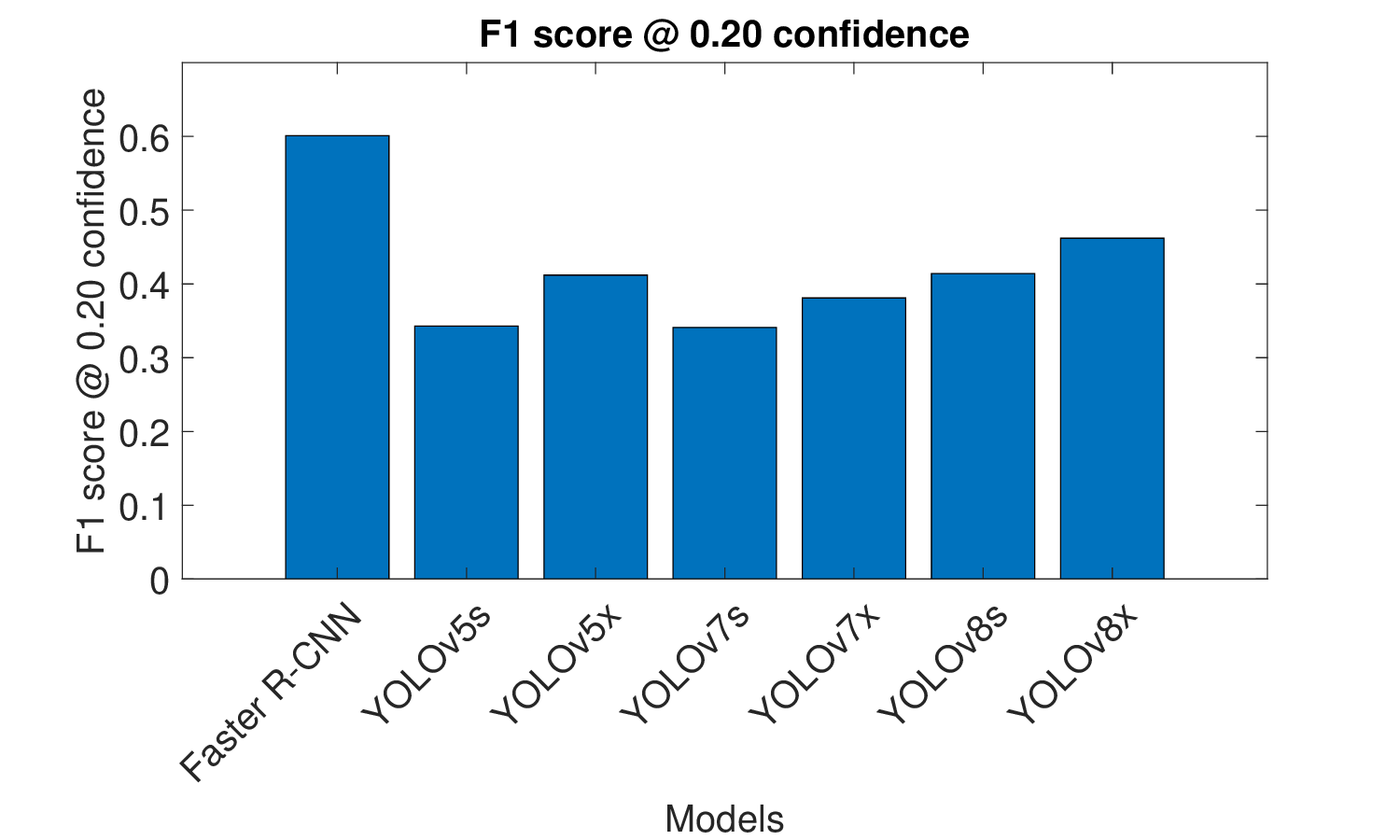}
    \caption{Comparison of F1 score (@0.2 confidence threshold).}
    \label{fig:f1score}
\end{subfigure}
\vspace{0.5cm}

\begin{subfigure}[b]{0.49\textwidth}
    \includegraphics[width=\textwidth]{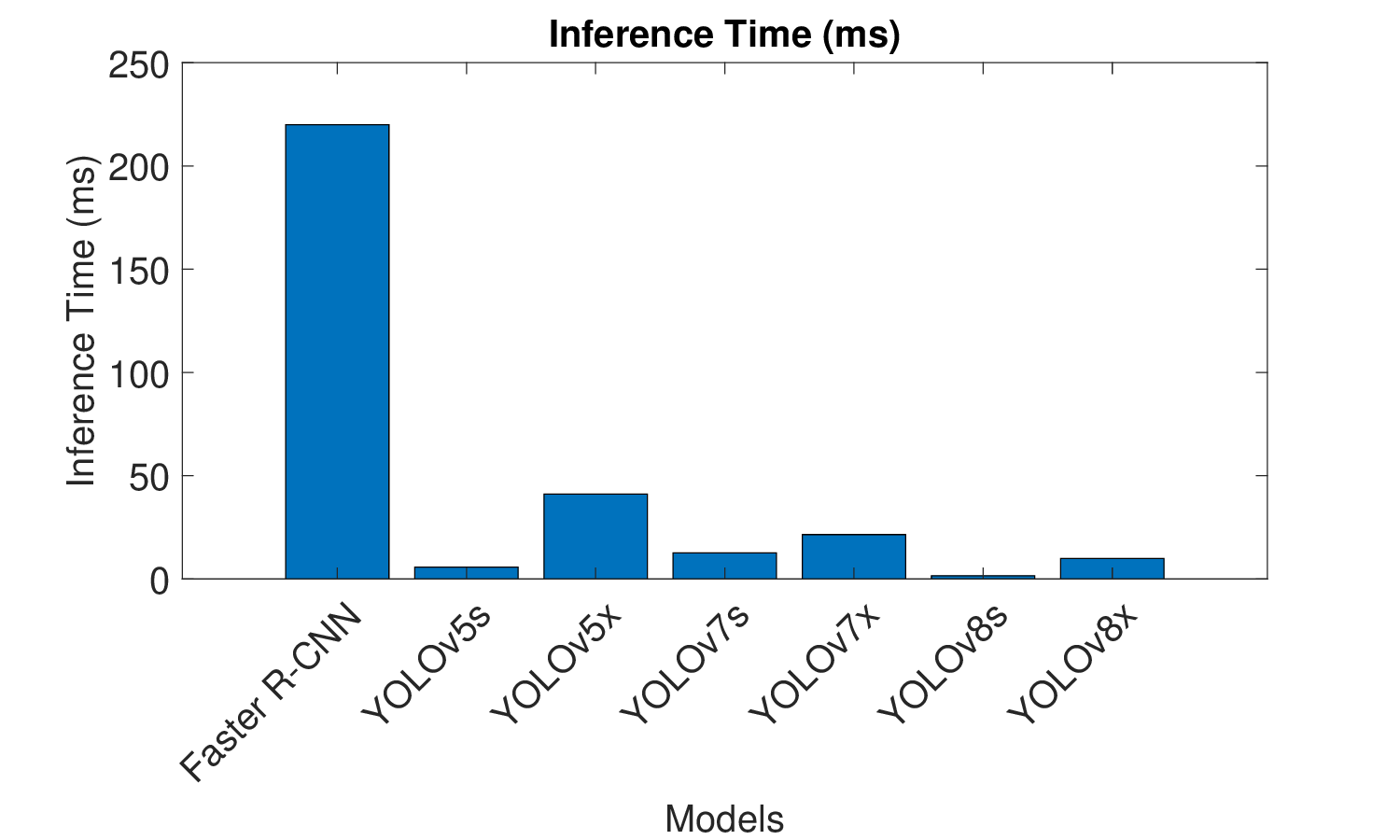}
    \caption{Comparison of inference time (milliseconds).}
    \label{fig:inference_time}
\end{subfigure}
\hfill
\begin{subfigure}[b]{0.49\textwidth}
    \includegraphics[width=\textwidth]{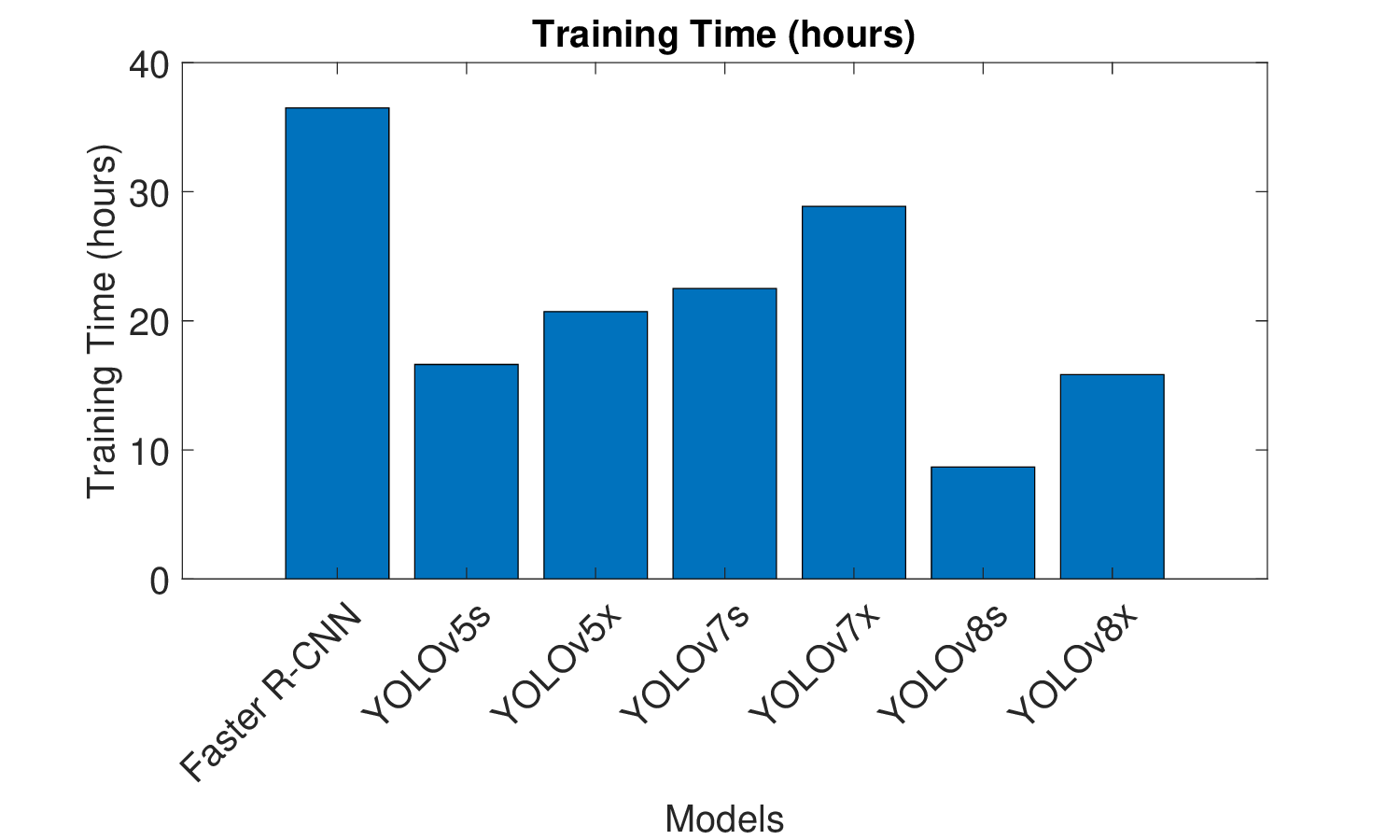}
    \caption{Comparison of training time (hours).}
    \label{fig:training_time}
\end{subfigure}

\caption{Performance comparison of evaluated frameworks in terms of precision, recall, mean average precision (mAP), F1 score, inference time, and training time.}
\label{fig:overall_comparison}
\end{figure}

The proposed YOLOv8 model was evaluated on the VisDrone2019 and Caltech Pedestrian datasets to assess its performance under various conditions. The YOLOv8x model outperformed all other models, including YOLOv5, YOLOv7, and Faster R-CNN, in detection metrics such as precision, recall, mAP @ 0.50, and F1 score. YOLOv8x achieved the highest precision (0.763), recall (0.485), mAP @ 0.50 (0.514), and F1 score (0.462) at a confidence threshold of 0.20, as shown in Figs. \ref{fig:precision}, \ref{fig:recall}, \ref{fig:mAP}, and \ref{fig:f1score}, respectively, and summarized in Table 1. YOLOv8s was the fastest model, having the lowest training time of all other models and achieving the highest FPS (625) and the lowest inference time (1.6 ms per image), but with an inference time of 9.9 ms per image and FPS of 101 as shown in Fig. \ref{fig:inference_time} and Fig. \ref{fig:training_time} respectively, YOLOv8x is also close in competition, making it overall most optimum option for real-time applications. In comparison, Faster R-CNN had significantly higher inference times (220 ms) and lower FPS (4.55), limiting its use in high-speed applications.

To evaluate the real-time performance of each detection algorithm, we estimated the computational time required to process 30 frames, which corresponds to 1 second of video at 30 frames per second (FPS). This was computed using the relation,

\[
\text{Computational Time (s)} = \frac{30}{\text{FPS}}.
\]

This metric provides an estimate of how much time each model takes to process one second of video input. It serves as a practical indicator of the algorithm's suitability for real-time VRU detection tasks, where lower computational time implies better real-time responsiveness.

\subsection{Class-wise Performance Comparison}
The comparison of the performance of our proposed model on two datasets VisDrone2019 and Caltech Pedestrian showed a stronger performance on VisDrone2019 for detecting bicycles and tricycles, while the Caltech Pedestrian dataset yielded better results for pedestrian and people detection, owning to larger number of instances for pedestrians and people and less number of instances for bicycle and tricycles. Table 2 provides a detailed breakdown of class-wise precision, recall, and mAP for both datasets. The comparison shows that overall mAP is 0.5415 with the VisDrone dataset, and with the Caltech Pedestrian dataset, mAP is 0.475, which is low. With this evaluation of achieving lower mAP, we set our ground for optimizing the gradient descent mechanism of the detection model to improve the mean average precision and avoid the catastrophic forgetting evident in the behavior of the existing YOLOv8 model.

\begin{sidewaystable}
\caption{Object detection evaluation metrics for YOLOv8x class-wise (pedestrian, people, tricycle, and bicycle) using VisDrone2019 dataset and Caltech Pedestrian dataset.}\label{tab:metrics}
\begin{tabular}{|l|l|l|l|l|l||l|l|l|l|l|}
\hline
\textbf{Class} & \textbf{Images} & \textbf{Instances} & \textbf{Precision} & \textbf{Recall} & \textbf{mAP50} & \textbf{Images} & \textbf{Instances} & \textbf{Precision} & \textbf{Recall} & \textbf{mAP50} \\
\hline
\multicolumn{6}{|c||}{\textbf{VisDrone}} & \multicolumn{5}{c|}{\textbf{Caltech Pedestrian}} \\
\hline
All & 2209 & 102083 & 0.763 & 0.485 & 0.5415 & 1500 & 159700 & 0.852 & 0.776 & 0.475 \\
\hline
Pedestrian & 2209 & 50600 & 0.664 & 0.560 & 0.556 & 1500 & 85000 & 0.950 & 0.950 & 0.950 \\
\hline
People & 2209 & 40230 & 0.640 & 0.500 & 0.556 & 1500 & 72000 & 0.750 & 0.600 & 0.675 \\
\hline
Tricycle & 2209 & 10933 & 0.553 & 0.328 & 0.362 & 1500 & 1500 & 0.120 & 0.060 & 0.081 \\
\hline
Bicycle & 2209 & 320 & 0.309 & 0.177 & 0.155 & 1500 & 1200 & 0.100 & 0.050 & 0.075 \\
\hline
\textbf{Overall mAP} & - & - & - & - & 0.5415 & - & - & - & - & 0.475 \\
\hline
\end{tabular}
\end{sidewaystable}

\begin{table}
\caption{Continual learning performance evaluation for dataset classes (pedestrian, people, tricycle, and bicycle) detected by YOLOv8x models. (a) Trained on VisDrone2019, (b) trained on Caltech Pedestrian dataset, (c) pre-trained on VisDrone then trained on Caltech with Adam optimizer, (d) pre-trained on VisDrone then trained on Caltech with SGD optimizer.}\label{tab3}
\begin{tabular}{|l|l|l|l|l|}
\hline
\textbf{Metric} & \textbf{YOLOv8x(a)} & \textbf{YOLOv8x(b)} & \textbf{YOLOv8x(c)} & \textbf{YOLOv8x(d)} \\
\hline
Precision @ 0.2 confidence & 0.763 & 0.852 & 0.855 & 0.889 \\
\hline
Recall @ 0.2 confidence & 0.485 & 0.776 & 0.772 & 0.783 \\
\hline
mAP @ 0.5 & 0.5415 & 0.475 & 0.461 & 0.608 \\
\hline
F1 score @ 0.2 confidence & 0.462 & 0.412 & 0.441 & 0.534 \\
\hline
Training time (hours) & 15.831 & 10.72 & 10.53 & 16.98 \\
\hline
\end{tabular}
\end{table}

\subsection{Continual Learning Evaluation}
When the YOLOv8x model was trained on the VisDrone2019 dataset and then fine-tuned on the Caltech Pedestrian dataset, the performance showed significant improvements. Table~3 column 1 shows the Precision (P), Recall (R) and F1 score of the detection model trained and tested on the VisDrone2019 dataset; however, Table~3 column 2 shows the Precision (P), Recall (R) and F1 score of detection model trained and tested from scratch on Caltech Pedestrian Dataset. Analysis of Column 2 shows that Precision (P), Recall (R), and F1 scores have improved when we train it from scratch on the Caltech Pedestrian Dataset as compared to the VisDrone2019 dataset because the size of objects of interest is large. Objects are in street view perspective in this dataset, and instances of two classes, namely pedestrians and people, are also large, as shown in Table~2. The decrease in mAP and F1 score is also evident, which is because for bicycles and tricycles instances, the number of instances is less in the Caltech Pedestrian Dataset as compared to the VisDrone dataset, resulting in an overall decrease in mean average precision (mAP) of all classes as shown in Table~2.

When the trained model on the VisDrone dataset is trained in Caltech Pedestrian Dataset with Adam optimizer Table ~3 column 3, the Precision (P), Recall (R), mAP and F1 score values are almost the same as achieved with from the scratch training on Caltech dataset Table ~3. This shows that the model has transferred its learning with negligible learning from the old training dataset VisDrone2019 and suffered from catastrophic forgetting. In order to handle the learning pattern of the model, we performed another iteration of model YOLOv8x training by changing the model optimizer to SGD. With this modification in the mechanism of converging the loss function, not only do we see improvement in Precision (P) and Recall (R) values but we also see a significant improvement in mAP and F1 score values Table ~3 column 4. This indicates that continual learning is performed successfully by changing the loss function optimization methodology. The training time for 300 epochs is shortest when the pre-trained YOLOv8x model on the VisDrone dataset is fine-tuned using the Adam optimizer, which is highly efficient. However, with the SGD optimizer, the training time increases significantly from 10.53 hours to 16.98 hours. This increase is likely due to the model's slower gradient optimization, as it works to retain the knowledge learned from the pre-trained weights.

YOLOv8x consistently outperformed other object detection models in terms of accuracy, with its high precision, recall, mAP, and F1 score, making it ideal for detecting VRUs in various real-world conditions. Continual learning improvements, such as switching to the SGD optimizer, enhanced the model's ability to learn from both datasets without catastrophic forgetting. The combination of high detection accuracy and real-time processing efficiency makes YOLOv8x a robust solution for road safety applications.

\section{Conclusion} \label{conclusion}
This study enhances the safety of SDVs, focusing on VRUs. We introduce a deep learning approach using the YOLOv8x model, which outperforms previous versions like YOLOv5 and YOLOv7 in accuracy and speed, especially in challenging environments. A key innovation is the continual learning mechanism in YOLOv8-D, allowing it to adapt to new datasets without forgetting prior knowledge. We aim to extend this work to develop a dynamic tracking system that supports SDVs by adapting to real-time requirements, paving the way for more resilient autonomous driving.

\textbf{Acknowledgements} The authors would like to thank Ministry of Higher Education Malaysia for Fundamental Research Grant Scheme with Project Code FRGS/1/2022/TK07/USM/02/14 for permitting them to carry out this research.

\bibliographystyle{splncs03_unsrt.bst}

\begin{thebibliography}{10}
\providecommand{\url}[1]{\texttt{#1}}
\providecommand{\urlprefix}{URL }

\bibitem{gatera2022towards}
GATERA, A.: Towards improved road traffic safety: A modelling and IoT integration approach. Ph.D. thesis, University of Rwanda (College of science and Technology) (2022)

\bibitem{WHO2018RoadSafety}
{World Health Organization}: Global status report on road safety 2018. Geneva (2018), licence: CC BY-NC-SA 3.0 IGO

\bibitem{WHO2023RoadSafety}
{World Health Organization}: Despite notable progress, road safety remains an urgent global issue (Dec 2023), \url{https://www.who.int/news/item/13-12-2023-despite-notable-progress-road-safety-remains-urgent-global-issue}, accessed: Accessed: 18-Jan-2024

\bibitem{UN_SDG2030}
{United Nations}: The 2030 agenda for sustainable development (2024), \url{https://sdgs.un.org/2030agenda}, accessed: 18-Jan-2024

\bibitem{djuric2020uncertainty}
Djuric, N., Radosavljevic, V., Cui, H., Nguyen, T., Chou, F.C., Lin, T.H., Singh, N., Schneider, J.: Uncertainty-aware short-term motion prediction of traffic actors for autonomous driving. In: Proceedings of the IEEE/CVF Winter Conference on Applications of Computer Vision. pp. 2095--2104 (2020)

\bibitem{constant2010protecting}
Constant, A., Lagarde, E.: Protecting vulnerable road users from injury. PLoS medicine  7(3),  e1000228 (2010)

\bibitem{elzein2003motion}
Elzein, H., Lakshmanan, S., Watta, P.: A motion and shape-based pedestrian detection algorithm. In: IEEE IV2003 Intelligent Vehicles Symposium. Proceedings (Cat. No. 03TH8683). pp. 500--504. IEEE (2003)

\bibitem{schauland2006vision}
Schauland, S., Kummert, A., Park, S.B., Iurgel, U., Zhang, Y.: Vision-based pedestrian detection--improvement and verification of feature extraction methods and svm-based classification. In: 2006 IEEE Intelligent Transportation Systems Conference. pp. 97--102. IEEE (2006)

\bibitem{xu2011pedestrian}
Xu, Y., Xu, L., Li, D., Wu, Y.: Pedestrian detection using background subtraction assisted support vector machine. In: 2011 11th International Conference on Intelligent Systems Design and Applications. pp. 837--842. IEEE (2011)

\bibitem{dimitrievski2019people}
Dimitrievski, M., Jacobs, L., Veelaert, P., Philips, W.: People tracking by cooperative fusion of radar and camera sensors. In: 2019 IEEE Intelligent Transportation Systems Conference (ITSC). pp. 509--514. IEEE (2019)

\bibitem{karakaya2023cyclesense}
Karakaya, A.S., Ritter, T., Biessmann, F., Bermbach, D.: Cyclesense: Detecting near miss incidents in bicycle traffic from mobile motion sensors. Pervasive and Mobile Computing  91,  101779 (2023)

\bibitem{islam2023traffic}
Islam, Z., Abdel-Aty, M.: Traffic conflict prediction using connected vehicle data. Analytic Methods in Accident Research  39,  100275 (2023)

\bibitem{redmon2016you}
Redmon, J., Divvala, S., Girshick, R., Farhadi, A.: You only look once: Unified, real-time object detection. In: Proceedings of the IEEE conference on computer vision and pattern recognition. pp. 779--788 (2016)

\bibitem{zhao2016faster}
Zhao, X., Li, W., Zhang, Y., Gulliver, T.A., Chang, S., Feng, Z.: A faster rcnn-based pedestrian detection system. In: 2016 IEEE 84th vehicular technology conference (VTC-Fall). pp. 1--5. IEEE (2016)

\bibitem{zhao2023yolov7}
Zhao, H., Zhang, H., Zhao, Y.: Yolov7-sea: Object detection of maritime uav images based on improved yolov7. In: Proceedings of the IEEE/CVF winter conference on applications of computer vision. pp. 233--238 (2023)

\bibitem{wang2023detector}
Wang, H., Jin, L., He, Y., Huo, Z., Wang, G., Sun, X.: Detector--tracker integration framework for autonomous vehicles pedestrian tracking. Remote Sensing  15(8),  2088 (2023)

\bibitem{tan2019efficientnet}
Tan, M., Le, Q.V.: Efficientnet: Rethinking model scaling for convolutional neural networks. arXiv preprint arXiv:1905.11946  (2019)

\bibitem{kirkpatrick2017overcoming}
Kirkpatrick, J., Pascanu, R., Rabinowitz, N., Veness, J., Desjardins, G., Rusu, A.A., Milan, K., Quan, J., Ramalho, T., Grabska-Barwinska, A., et~al.: Overcoming catastrophic forgetting in neural networks. Proceedings of the national academy of sciences  114(13),  3521--3526 (2017)

\bibitem{torrey2010transfer}
Torrey, L., Shavlik, J.: Transfer learning. In: Handbook of research on machine learning applications and trends: algorithms, methods, and techniques, pp. 242--264. IGI global (2010)

\bibitem{weiss2016survey}
Weiss, K., Khoshgoftaar, T.M., Wang, D.: A survey of transfer learning. Journal of Big data  3(1),  1--40 (2016)

\bibitem{hassabis2017neuroscience}
Hassabis, D., Kumaran, D., Summerfield, C., Botvinick, M.: Neuroscience-inspired artificial intelligence. Neuron  95(2),  245--258 (2017)

\bibitem{arpit2017closer}
Arpit, D., Jastrz{\k{e}}bski, S., Ballas, N., Krueger, D., Bengio, E., Kanwal, M.S., Maharaj, T., Fischer, A., Courville, A., Bengio, Y., et~al.: A closer look at memorization in deep networks. In: International conference on machine learning. pp. 233--242. PMLR (2017)

\bibitem{lopez2017gradient}
Lopez-Paz, D., Ranzato, M.: Gradient episodic memory for continual learning. Advances in neural information processing systems  30 (2017)

\bibitem{zenke2017continual}
Zenke, F., Poole, B., Ganguli, S.: Continual learning through synaptic intelligence. In: International conference on machine learning. pp. 3987--3995. PMLR (2017)

\bibitem{aljundi2018memory}
Aljundi, R., Babiloni, F., Elhoseiny, M., Rohrbach, M., Tuytelaars, T.: Memory aware synapses: Learning what (not) to forget. In: Proceedings of the European conference on computer vision (ECCV). pp. 139--154 (2018)

\bibitem{hinton2015distilling}
Hinton, G., Vinyals, O., Dean, J.: Distilling the knowledge in a neural network. arXiv preprint arXiv:1503.02531  (2015)

\bibitem{li2017learning}
Li, Z., Hoiem, D.: Learning without forgetting. IEEE transactions on pattern analysis and machine intelligence  40(12),  2935--2947 (2017)

\bibitem{zhu2021detection}
Zhu, P., Wen, L., Du, D., Bian, X., Fan, H., Hu, Q., Ling, H.: Detection and tracking meet drones challenge. IEEE Transactions on Pattern Analysis and Machine Intelligence  44(11),  7380--7399 (2021)

\bibitem{dollar_wojek_schiele_perona_2009}
Dollar, P., Wojek, C., Schiele, B., Perona, P.: Caltech pedestrians (Jun 2009)

\end{thebibliography}

\end{document}